\documentclass[conference]{IEEEtran}
\IEEEoverridecommandlockouts
\usepackage[style=ieee,url=false,doi=false,isbn=false]{biblatex}
\usepackage{amsmath,amssymb,amsfonts}
\usepackage{graphicx}
\usepackage{tikz}
\usetikzlibrary{quantikz2}
\usepackage{textcomp}
\usepackage{xcolor}
\usepackage{braket}
\usepackage{physics}
\usepackage{subfigure}
\usepackage{algorithm}
\usepackage{algpseudocode}
\usepackage{bm}
\usepackage{booktabs}
\def\BibTeX{{\rm B\kern-.05em{\sc i\kern-.025em b}\kern-.08em
    T\kern-.1667em\lower.7ex\hbox{E}\kern-.125emX}}

\definecolor{red}{HTML}{f44336}
\definecolor{thinred}{HTML}{E57373}
\definecolor{orange}{HTML}{FF9800}
\definecolor{yellow}{HTML}{FFEB3B}
\definecolor{green}{HTML}{4CAF50}
\definecolor{blue}{HTML}{2196F3}
\definecolor{indigo}{HTML}{3F51B5}
\definecolor{purple}{HTML}{9C27B0}
\definecolor{gray}{HTML}{9E9E9E}

\begin{document}

\title{
  Quantum hardware noise learning via differentiable Kraus representation on tensor networks
  \thanks{
    This paper is based on results obtained from a project, JPNP25014, commissioned by the New Energy and Industrial Technology Development Organization (NEDO).
  }
}

\makeatletter
\newcommand{\linebreakand}{%
\end{@IEEEauthorhalign}
\hfill\mbox{}\par
\mbox{}\hfill\begin{@IEEEauthorhalign}
}
\makeatother

\author{
  \IEEEauthorblockN{
    Ryo Sakai and Yu Yamashiro
  }
  \IEEEauthorblockA{
    \textit{JIJ Inc.}, 3-3-6 Shibaura, Minato-ku, Tokyo, 108-0023, Japan
  }
  \IEEEauthorblockA{
    r.sakai@j-ij.com, y.yamashiro@j-ij.com
  }
}

\maketitle

\begin{abstract}
  We present a method for learning quantum hardware noise from a measurement distribution of a single device experiment.
  Each noise channel is represented by automatically differentiable Kraus operators obtained from a Stinespring-based parameterization that is completely positive and trace preserving by construction,
  and circuits are simulated with a matrix product density operator forward model.
  Independent channels are attached to each native gate type, to each nearest-neighbor crosstalk interaction, and to state preparation and measurement,
  and all channels are optimized end-to-end against a distance between the simulated and observed measurement distributions.
  On ibm\_fez, a Heron-generation superconducting processor, training on a ripple-carry adder circuit reproduces the device output distribution,
  and the same learned parameters, applied without retraining, also track the device distribution of an unrelated multiplier circuit,
  indicating that the method captures intrinsic device characteristics rather than overfitting to the training circuit.
  A systematic evaluation across a range of benchmark circuits confirms that this generalization is consistent.
  We further use the learned model to perform an offline feasibility assessment of the quantum approximate optimization algorithm with an error detection scheme,
  demonstrating the kind of noise-aware prediction the framework is designed to enable.
\end{abstract}


\section{Introduction}
\label{sec:introduction}

Noise remains the dominant obstacle to extracting useful computation from current quantum hardware,
and every error-mitigation, circuit-optimization, or benchmarking technique that aspires to correct for it starts from the same prerequisite:
a faithful model of what the device actually does.
Vendor-provided calibration data reports per-component figures of merit such as gate infidelities, but does not provide a unified, device-wide noise model from which the output distribution of an arbitrary circuit can be predicted.
Traditional quantum process tomography~\cite{Chuang:1996hw} is accurate but expensive, requiring informationally complete probe states and measurement settings,
and it characterizes one gate at a time rather than the device as a whole.
These limitations motivate a different set of requirements: a noise model that is expressive enough to represent general completely positive and trace-preserving (CPTP) maps,
device-wide so that it covers every gate, crosstalk interaction, and SPAM channel simultaneously,
trainable from ordinary circuit-execution data rather than bespoke tomographic protocols,
and transferable from the training circuit to new algorithms of interest without retraining.

Existing approaches fall short of these requirements in different ways.
A prominent direction uses tensor networks to learn noise from measurement data.
Torlai \textit{et al.}~\cite{Torlai:2020fgl} represented a quantum process as a tensor network and optimized its tensor elements by gradient descent against tomographic data,
achieving process fidelities above $0.99$ on circuits of up to 10 qubits.
Mangini \textit{et al.}~\cite{Mangini:2024jyb} extended this to characterize correlated noise channels affecting entire gate layers on systems of up to 20 qubits, combining it with an error mitigation scheme.
In both cases, the variational parameters are the tensor elements of a process-level representation---the full quantum channel is learned as a monolithic tensor network, without decomposition into per-gate noise channels.
This gives an accurate description of the overall process but does not yield a gate-level noise model that can be transferred to other circuits.
Ma \textit{et al.}~\cite{Ma:2024deb} took a complementary route:
they performed traditional quantum process tomography on individual gates, extracted Kraus operators via Choi-matrix eigendecomposition, and inserted them into a tensor network simulator.
This produces per-gate channels, but the tomography and simulation stages are decoupled---no gradient flows from the circuit-level distribution mismatch back to the channel parameters.
Filippov \textit{et al.}~\cite{Filippov:2022exc} introduced the matrix product channel for post-processing noisy variational quantum eigensolver outputs to mitigate both noise- and ansatz-related errors;
while the optimization is end-to-end, the learned object is a correction channel rather than a device noise model.
All four approaches rely on tomographic probe states or informationally complete measurements rather than the output distribution of an arbitrary circuit.

Several approaches parameterize quantum channels directly and learn them.
Differentiable process tomography that learns Kraus operators on the Stiefel manifold~\cite{Ahmed:2022ioj} or via Stinespring dilation~\cite{Visser:2023qiq} enables gradient-based channel learning,
but targets a single quantum channel from tomographic probe data rather than a device-wide architecture spanning gate, crosstalk, and SPAM noise.
Error-correction-oriented noise estimation via Bayesian inference or differentiable maximum likelihood on syndrome statistics~\cite{Kobori:2024loi, Cao:2026hqb} recovers noise parameters for decoders in surface or repetition codes rather than full CPTP maps on general circuits.
Non-Markovian process tensor tomography and structured model identification~\cite{White:2023kpy, Jamadagni:2026bht, Ji:2025toi} extract temporally correlated or physically motivated noise at scale,
but either require bespoke experimental protocols or trade model expressivity for identifiability.
Complementary work on tensor-network state tomography and shadow-based channel reconstruction~\cite{Guo:2023txh, Votto:2025zah, Liu:2026pzn} learns states and channels from local or randomized measurements rather than gate-level device noise.
To our knowledge, no existing method combines a fully general, device-wide, differentiable-Kraus noise architecture trained end-to-end from the computational-basis output distribution of a single hardware circuit,
with a demonstration that the learned model transfers across circuits on real hardware.

In this work we present such a method, built on several ingredients.
We parameterize each noise channel via a unitary $U(\theta) = e^{iH(\theta)}$ acting on the system-plus-ancilla Hilbert space and extract Kraus operators as $K_k = \bra{k}_{\mathrm{anc}} U(\theta) \ket{0}_{\mathrm{anc}}$,
guaranteeing the trace-preserving condition $\sum_k K_k^\dagger K_k = I$ by construction.
Independent channels are attached to every native gate type, to each nearest-neighbor crosstalk interaction, and to state preparation and measurement.
Noisy circuits are simulated using matrix product density operators (MPDO)~\cite{Verstraete:2004gdw,2021PhRvR...3b3005C,Guo:2023dqn},
a tensor-network ansatz that represents mixed states by attaching an inner (Kraus) index to each physical site and absorbs noise channels by expanding that index followed by singular value decomposition (SVD) with truncation.
The entire pipeline is end-to-end differentiable via automatic differentiation through the SVD and QR decompositions~\cite{Liao:2019bye}, enabling gradient computation in JAX~\cite{jax}.
As the measure of distributional distance, we consider two loss functions:
a negative log-likelihood for direct maximum-likelihood estimation,
and an entropy-regularized optimal transport (OT) with a Hamming ground metric, solved by a Sinkhorn iteration~\cite{cuturi2013sinkhorn}.
All channels are optimized jointly against a single experiment on real hardware.

We validate the method on both synthetic and hardware data.
On the 10-qubit QASMBench~\cite{Li:2020fmi} adder\_n10~\cite{Cuccaro:2004xxx} circuit with injected single-qubit bit-flip and two-qubit depolarizing noise,
the learned channels recover the ground-truth Kraus operators with Choi process fidelity above $0.99$,
including the two-qubit $CZ$ channel that involves entangling noise between sites.
On ibm\_fez, an IBM's Heron-generation superconducting processor,
training on a hardware measurement distribution of the same adder\_n10 reproduces the long-tailed 32-outcome distribution of the device,
including the dominant correct output $\ket{10000}$ and the ranking of the remaining bitstrings.
Most importantly, the same learned noise parameters, applied without retraining,
also track the device distribution of an unrelated 13-qubit multiply\_n13 circuit,
indicating that the model captures intrinsic device characteristics rather than overfitting to the training circuit.
A systematic evaluation across a range of QASMBench circuits, comparing the learned model against random CPTP noise of matched magnitude, confirms that this advantage is consistent rather than circuit-specific.
We further use the learned model to perform an offline, circuit-level feasibility assessment of parity-check-error-detected QAOA~\cite{Shaydulin:2023fpr} for the low autocorrelation binary sequences (LABS) problem~\cite{1053969,1054411},
a concrete example of the noise-aware algorithmic prediction the framework is designed to enable.

This paper is organized as follows.
Section~\ref{sec:method} describes the noise channel formalism and its Stinespring parameterization,
the device-wide noise model architecture covering gate noise, crosstalk, and SPAM errors,
the automatically differentiable MPDO simulator,
and the training procedure with NLL and OT losses.
Section~\ref{sec:results} presents results on both synthetic and ibm\_fez hardware noise,
including per-channel process fidelities, the cross-circuit generalization test, and the LABS-QAOA feasibility assessment.
Section~\ref{sec:summary} summarizes our findings, discusses connections to related work, and outlines directions for future work.

\section{Method}
\label{sec:method}

\subsection{Quantum noise channels}

A quantum noise channel $\mathcal{E}$ acting on a density matrix $\rho$ is a CPTP map that can be written in the Kraus representation~\cite{kraus1983states} as
\begin{align}
  \mathcal{E}(\rho) = \sum_{k=0}^{N_{\mathrm{K}}-1} K_k \rho K_k^\dagger,
  \label{eq:kraus}
\end{align}
where $N_{\mathrm{K}}$ is the number of Kraus operators and $\{K_k\}$ satisfy the
completeness relation $\sum_k K_k^\dagger K_k = I$.
For a single-qubit channel, each $K_k \in \mathbb{C}^{2\times 2}$,
while for a two-qubit channel, $K_k \in \mathbb{C}^{4\times 4}$.

Standard noise channels such as depolarizing, dephasing, amplitude damping, and Pauli channels are all special cases of eq.~\eqref{eq:kraus} with specific structures imposed on $\{K_k\}$.
In this work, we do not assume any such structure---instead, we learn general Kraus operators from data.

\subsection{Stinespring parameterization of Kraus operators}

The Stinespring dilation theorem~\cite{Stinespring:1955eig} guarantees that every CPTP map on a $d$-dimensional system can be realized as a unitary on a larger system--ancilla Hilbert space $\mathcal{H}_S \otimes \mathcal{H}_A$, where $\dim \mathcal{H}_S = d$ and $\dim \mathcal{H}_A = N_{\mathrm{K}}$, followed by tracing out the ancilla.
Concretely, for any unitary $U \in \mathbb{C}^{dN_{\mathrm{K}} \times dN_{\mathrm{K}}}$, a fixed ancilla reference state $\ket{0}_A$, and an arbitrary system state $\rho$, the map
\begin{align}
  \mathcal{E}(\rho)
  = \mathrm{Tr}_A \left[ U (\rho \otimes \dyad{0}_A) U^\dagger \right]
  = \sum_{k=0}^{N_{\mathrm{K}}-1} K_k \rho K_k^\dagger
\end{align}
is CPTP, where the Kraus operators are the $d \times d$ blocks
\begin{align}
  \label{eq:kraus_extract}
  K_k = \bra{k}_A U \ket{0}_A = U_{[kd:(k+1)d, 0:d]}.
\end{align}
The completeness relation $\sum_k K_k^\dagger K_k = I_d$ follows directly from the unitarity of $U$:
\begin{align}
  \sum_k K_k^\dagger K_k
  = \bra{0}_A U^\dagger
  \left(\sum_k \dyad{k}_A\right)
  U \ket{0}_A
  = I_d.
\end{align}

A general CPTP map on a $d$-dimensional system requires at most $N_{\mathrm{K}} = d^2$ Kraus operators~\cite{Choi:1975nug}.
For a single qubit ($d=2$), this means $N_{\mathrm{K}}^{\max} = 4$, so any single-qubit noise channel---no matter how complex---can be represented exactly with four Kraus operators.
For two qubits ($d=4$), the maximum is $N_{\mathrm{K}}^{\max} = 16$.
In our experiments we use $N_{\mathrm{K}} = 4$ for all channels;
this is exact for single-qubit channels and approximation for two-qubit channels, providing a practical trade-off between expressiveness and numerical complexity.

We exploit the Stinespring structure to build a differentiable parameterization of Kraus operators.
Given a real parameter vector $\theta \in \mathbb{R}^{(dN_{\mathrm{K}})^2}$, we construct a Hermitian matrix $H(\theta) \in \mathbb{C}^{dN_{\mathrm{K}} \times dN_{\mathrm{K}}}$ whose diagonal entries are the first $dN_{\mathrm{K}}$ components of $\theta$ and whose upper-triangular real and imaginary parts are filled by the remaining components,
ensuring $H = H^\dagger$ by construction.
The unitary is then $U(\theta) = e^{i H(\theta)}$.
Since every Hermitian matrix in $\mathbb{C}^{dN_{\mathrm{K}} \times dN_{\mathrm{K}}}$ is reachable by some $\theta$, and the matrix exponential maps Hermitian matrices surjectively onto the unitary group $U(dN_{\mathrm{K}})$, this parameterization can represent any CPTP map with at most $N_{\mathrm{K}}$ Kraus operators.
Moreover, the mapping $\theta \mapsto U(\theta)$ is differentiable with respect to $\theta$, and so are the Kraus operators $K_k(\theta)$ since they are obtained from $U(\theta)$ by block slicing~\eqref{eq:kraus_extract}.

The complete procedure is summarized in algorithm~\ref{alg:kraus}.

\begin{algorithm}[H]
  \caption{Kraus operators from parameters}
  \label{alg:kraus}
  \begin{algorithmic}[1]
    \Require Parameter vector $\theta \in \mathbb{R}^{(dN_{\mathrm{K}})^2}$, system dimension $d$, number of Kraus operators $N_{\mathrm{K}}$
    \Ensure Kraus operators $\{K_0, \ldots, K_{N_{\mathrm{K}}-1}\}$ satisfying $\sum_k K_k^\dagger K_k = I_d$
    \State $n \leftarrow d N_{\mathrm{K}}$
    \State $D \leftarrow \mathrm{diag}(\theta_1, \ldots, \theta_n)$ \Comment{real diagonal}
    \State $U_{\mathrm{tri}} \leftarrow$ upper-triangular matrix with entries $\theta_{n+2j-1} + i \theta_{n+2j}$ for $\displaystyle j = 1, \ldots, n$
    \State $H \leftarrow D + U_{\mathrm{tri}} + U_{\mathrm{tri}}^\dagger$ \Comment{Hermitian by construction}
    \State $U \leftarrow e^{i H}$ \Comment{unitary by exponentiation}
    \For{$k = 0, \ldots, N_{\mathrm{K}}-1$}
      \State $K_k \leftarrow U_{[kd:(k+1)d,\; 0:d]}$ \Comment{block slicing}
    \EndFor
  \end{algorithmic}
\end{algorithm}

For single-qubit channels ($d=2$, $N_{\mathrm{K}}=4$) the parameter vector has $(2 \times 4)^2 = 64$ real entries.
For two-qubit channels ($d=4$, $N_{\mathrm{K}}=4$) it has $(4 \times 4)^2 = 256$ entries.

Several alternative differentiable parameterizations of CPTP maps have been studied in the literature.
The earliest line of work parameterizes Kraus operators directly as points on the Stiefel manifold of isometries.
This parameterization has been used to analyze control landscapes for open quantum systems~\cite{Pechen:2008key,Oza:2009wvn}.
It also underlies gradient-descent quantum process tomography via constrained Kraus-operator learning~\cite{Ahmed:2022ioj} and the more general Riemannian-optimization and automatic-differentiation framework of Luchnikov \textit{et al.}~\cite{Luchnikov:2020eic}.
A very recent alternative in ref.~\cite{Ateeq:2025oia} encodes Kraus operators as unit vectors on a hypersphere with orthogonality and symplecticity constraints, giving a manifestly CPTP and differentiable parameterization without relying on ancilla dilation.
The closest prior work to our channel-side parameterization is ref.~\cite{Visser:2023qiq}, which uses the Stinespring dilation theorem to variationally learn the dilation unitary of a single target channel on a neutral-atom platform, in either a gate- or pulse-based ansatz, from measurement data;
by contrast, in our current work, the Stinespring-based parameterization is embedded in a device-wide noise-model architecture of many independent channels and trained end-to-end through an MPDO forward model as seen below.

\subsection{Noise model architecture}
\label{sec:noise_arch}

We assign independent learnable (\textit{i.e.} automatically differentiable) Kraus operators to each type of operation in the quantum circuit.
A state-preparation noise channel is applied to each qubit after initialization in $\ket{0}$.
A separate single-qubit noise channel is assigned to each native single-qubit gate type,
and a separate two-qubit noise channel is assigned to each native two-qubit gate type.
To capture nearest-neighbor crosstalk noise, a single-qubit channel is applied to all neighbors of the gate qubit(s) (as defined by the device topology) whenever a gate is executed.
Finally, a measurement-noise channel acts on each qubit immediately before readout.

We note that preparation, measurement, and crosstalk channels are all modeled as independent single-qubit channels in this work.
In practice, these processes can exhibit multi-qubit correlations---\textit{e.g.} readout crosstalk between neighboring qubits or joint system-neighbor dynamics during a gate.
Capturing such correlations would require multi-qubit channels for these operations, at the cost of a larger parameter count and an increased risk of overfitting.
We leave such extensions to future work and adopt the single-qubit assumption throughout as a practical trade-off.

Figure~\ref{fig:noise_model} illustrates the layout of the noise channels on a linearly connected four-qubit example.
Each ideal operation in the circuit is followed by its associated noise channel:
a one-qubit channel after each single-qubit gate with crosstalk channel on each nearest neighbor,
and a two-qubit channel after each two-qubit gate together with crosstalk on the neighbors of both gate qubits.
State preparation and measurement are also wrapped in their own one-qubit channels.

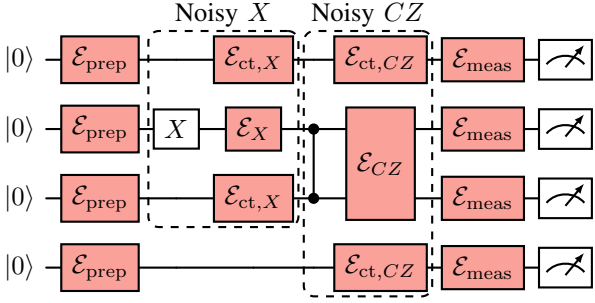
\begin{figure}[htbp]
  \centering
  \tikzset{ng/.style={fill=red!50}}
  \begin{quantikz}[column sep=0.2cm, row sep=0.3cm]
    \lstick{$\ket{0}$} & \gate[style=ng]{\mathcal{E}_\mathrm{prep}} & \gategroup[wires=3,steps=2,style={dashed,rounded corners,inner sep=-1.5pt}]{Noisy $X$} & \gate[style=ng]{\mathcal{E}_{\mathrm{ct},X}} & \gategroup[wires=4,steps=2,style={dashed,rounded corners,inner sep=-1.5pt}]{Noisy $CZ$} & \gate[style=ng]{\mathcal{E}_{\mathrm{ct},CZ}} & \gate[style=ng]{\mathcal{E}_\mathrm{meas}} & \meter{} \\
    \lstick{$\ket{0}$} & \gate[style=ng]{\mathcal{E}_\mathrm{prep}} & \gate{X} & \gate[style=ng]{\mathcal{E}_{X}} & \ctrl{1} & \gate[2,style=ng]{\mathcal{E}_{CZ}} & \gate[style=ng]{\mathcal{E}_\mathrm{meas}} & \meter{} \\
    \lstick{$\ket{0}$} & \gate[style=ng]{\mathcal{E}_\mathrm{prep}} & & \gate[style=ng]{\mathcal{E}_{\mathrm{ct},X}} & \control{} & & \gate[style=ng]{\mathcal{E}_\mathrm{meas}} & \meter{} \\
    \lstick{$\ket{0}$} & \gate[style=ng]{\mathcal{E}_\mathrm{prep}} & & & & \gate[style=ng]{\mathcal{E}_{\mathrm{ct},CZ}} & \gate[style=ng]{\mathcal{E}_\mathrm{meas}} & \meter{}
  \end{quantikz}
  \caption{
    Noise model architecture on a linear four-qubit chain.
    Red boxes denote Kraus channels.
    Each qubit is wrapped in independent state-preparation ($\mathcal{E}_\mathrm{prep}$) and measurement ($\mathcal{E}_\mathrm{meas}$) channels.
    After each single-qubit gate (here $X$ on the second qubit) the gate qubit receives a single-qubit gate-noise channel ($\mathcal{E}_{X}$) and every nearest neighbor receives a crosstalk channel ($\mathcal{E}_{\mathrm{ct},X}$).
    After each two-qubit gate ($CZ$ on the second and third qubits) a two-qubit channel ($\mathcal{E}_{CZ}$) is applied to the gate qubits and a crosstalk channel ($\mathcal{E}_{\mathrm{ct},CZ}$) acts on every neighbor that is not part of the gate.
    A linear chain is used here for illustration only; in our experiments the crosstalk neighbors are determined by the heavy-hex connectivity of the IBM Heron-generation processors.
  }
  \label{fig:noise_model}
\end{figure}

\subsection{Matrix product density operators}
\label{sec:mpdo}

A matrix product density operator (MPDO), also known as a locally purified density operator (LPDO)~\cite{Verstraete:2004gdw}, represents the density matrix of an $N$-qubit system as a tensor network.
Since applying noise channels to a pure state generically produces a mixed state, a density-operator representation is essential in this work~\cite{2021PhRvR...3b3005C,Guo:2023dqn}.
For each site $i$, the MPDO tensor has four indices: $A^{[i]}_{\alpha_{i-1}, s_i, a_i, \alpha_i}$,
where $\alpha_{i-1}$, $\alpha_i$ are left and right virtual indices (dimension $\chi$),
$s_i$ is the physical index (dimension 2 for a qubit),
and $a_i$ is the inner (Kraus) index (dimension $\kappa$).
The density matrix is recovered as
\begin{align}
  \rho_{s_1 \cdots s_N, s_1' \cdots s_N'}
  = \sum_{\{\alpha_i, \alpha_i', a_i\}} \prod_i
  A^{[i]}_{\alpha_{i-1}, s_i, a_i, \alpha_i} \; A^{[i]*}_{\alpha_{i-1}', s_i', a_i, \alpha_i'}.
\end{align}
Although some works (see \textit{e.g.} ref.~\cite{Lee:2025jpe}) simulate noisy circuits by representing the density operator directly as an MPO without an inner dimension,
bond truncation in this setting can violate Hermiticity, yielding unphysical density matrices.
The MPDO formalism with the inner indices guarantees Hermiticity by construction.

The MPDO exploits a structural analogy between quantum entanglement and classical mixture:
entanglement between sites is captured by the virtual indices,
while the classical mixing introduced by noise channels is captured by the inner indices.
Separating these two sources of complexity into independent indices allows each to be truncated independently.
Generally, selecting a sufficient bond dimension is not a trivial task, and the same applies to the inner dimension.
Throughout this paper, we set the inner dimension as $\kappa = 2 \chi$ following the prescription in ref.~\cite{2021PhRvR...3b3005C}.
We draw the reader's attention to ref.~\cite{Wei:2026rae}, which showed that, for certain classes of circuits, noisy dynamics exponentially contract truncation errors, suggesting a favorable error scaling for tensor-network-based simulation in general.

Given a circuit and the noise model of sec.~\ref{sec:noise_arch}, the full simulation applies each gate and its associated noise channels sequentially on the MPDO tensors,
producing a final state $\rho(\theta)$ from which measurement probabilities and samples can be extracted.

For a single-site noise channel with Kraus operators $\{K_k\}_{k=0}^{N_{\mathrm{K}}-1}$ acting on site~$i$, the MPDO tensor $A^{[i]}_{\alpha, s, a, \beta}$ is updated in three steps.
First, each Kraus operator is contracted with the physical index:
$\tilde{A}^{[i](k)}_{\alpha, s', a, \beta} = \sum_s (K_k)_{s',s} A^{[i]}_{\alpha, s, a, \beta}$.
The resulting $N_{\mathrm{K}}$ tensors are then stacked along the inner dimension to form $\hat{A}^{[i]}_{\alpha, s, (k,a), \beta}$ with new inner dimension $N_{\mathrm{K}} \kappa$.
Finally, the inner dimension is SVD-truncated back to $\kappa$ to control computational complexity.

For a two-site noise channel with $N_{\mathrm{K}}$ Kraus operators $\{K_k\} \in \mathbb{C}^{4 \times 4}$ acting on adjacent sites $i$ and $i+1$, the procedure is illustrated in fig.~\ref{fig:two_site_noise}.
The two MPDO tensors $A^{[i]}$ and $A^{[i+1]}$ are first merged by contracting their shared virtual index.
Each Kraus operator $K_k$ is applied to the joint physical index $(s_i, s_{i+1})$.
The Kraus index $k$ is then factored as $k \to (k_1, k_2)$ with $k_1, k_2 \in \{0, \ldots, \lceil \sqrt{N_{\mathrm{K}}} \rceil - 1\}$,
distributing the noise symmetrically between the two sites:
$k_1$ is absorbed into site~$i$'s inner dimension and $k_2$ into site~$i+1$'s,
giving each site a temporary inner dimension $\lceil \sqrt{N_{\mathrm{K}}} \rceil \kappa$.
The merged tensor is then SVD-decomposed back into two site tensors truncating the bond dimension to $\chi$,
and each site's inner dimension is separately SVD-truncated back to $\kappa$.

\begin{figure}[htbp]
  \centering
  \begin{tikzpicture}[
    tensor/.style={draw, fill=indigo!50, minimum size=0pt, font=\scriptsize},
    lbl/.style={font=\scriptsize},
    scale=0.8
    ]
    \node[tensor] (Ai) at (0, 0) {$A^{[1]}$};
    \node[tensor] (Aj) at (1, 0) {$A^{[2]}$};
    \draw (Ai.west)  -- ++(-0.4,0);
    \draw (Ai.east)  -- (Aj.west);
    \draw (Aj.east)  -- ++(0.4,0);
    \draw (Ai.north) -- ++(0,0.2);
    \draw (Aj.north) -- ++(0,0.2);
    \draw (Ai.south) -- ++(0,-0.2) node[below,lbl] {$\kappa$};
    \draw (Aj.south) -- ++(0,-0.2) node[below,lbl] {$\kappa$};

    \draw[->] (2,0) -- (3,0)
    node[midway,above,align=center,font=\scriptsize] {merge \\ + \\ stack Kraus effects \\};

    \node[tensor, minimum width=1.5cm] (M) at (4.5, 0) {$M$};
    \draw (M.west)  -- ++(-0.4,0);
    \draw (M.east)  -- ++(0.4,0);
    \draw ([xshift=-8pt]M.north) -- ++(0,0.2);
    \draw ([xshift=8pt]M.north)  -- ++(0,0.2);
    \draw[ultra thick] ([xshift=-8pt]M.south) -- ++(0,-0.2)
    node[below,lbl,xshift=-12pt] {$\lceil \sqrt{N_{\mathrm{K}}} \rceil \kappa$};
    \draw[ultra thick] ([xshift=8pt]M.south)  -- ++(0,-0.2)
    node[below,lbl,xshift=12pt] {$\lceil \sqrt{N_{\mathrm{K}}} \rceil \kappa$};

    \draw[->] (6.0,0) -- (7,0)
    node[midway,above,align=center,font=\scriptsize] {SVDs \\};

    \node[tensor] (Bi) at (8, 0) {$\tilde{A}^{[1]}$};
    \node[tensor] (Bj) at (9, 0) {$\tilde{A}^{[2]}$};
    \draw (Bi.west)  -- ++(-0.4,0);
    \draw (Bi.east)  -- (Bj.west);
    \draw (Bj.east)  -- ++(0.4,0);
    \draw (Bi.north) -- ++(0,0.2);
    \draw (Bj.north) -- ++(0,0.2);
    \draw (Bi.south) -- ++(0,-0.2) node[below,lbl] {$\kappa$};
    \draw (Bj.south) -- ++(0,-0.2) node[below,lbl] {$\kappa$};
  \end{tikzpicture}
  \caption{
    Two-site noise application.
    Inner dimensions (bottom legs) expand from $\kappa$ to $\lceil \sqrt{N_{\mathrm{K}}} \rceil \kappa$ after Kraus application,
    then are truncated back to $\kappa$.
  }
  \label{fig:two_site_noise}
\end{figure}
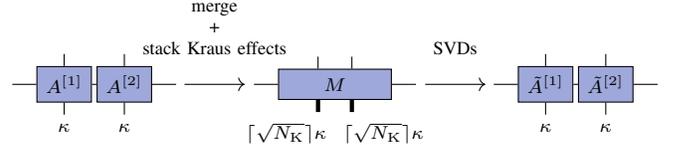

The probability of observing a bitstring $x = (x_1, \ldots, x_N)$ is $p_x = \mathrm{Tr}[\rho \dyad{x}]$.
In the MPDO this reduces to a sequential contraction of tensors in which the physical index of each tensor is projected onto $x_i$ and the inner indices are summed out.
This contraction is fully differentiable and provides the forward model through which the loss functions in sec.~\ref{sec:training} are evaluated.

\subsection{Training on distribution from hardware}
\label{sec:training}

Our goal is to find the noise parameters $\theta$ such that the measurement distribution of the MPDO simulation matches the empirical distribution observed on quantum hardware.
We formulate this as a minimization problem over a loss function that quantifies the discrepancy between the two distributions, and solve it via gradient-based optimization.

Given observed bitstrings $\{x_i\}$ with counts $\{n_i\}$ from $N_{\mathrm{shots}}$ total shots, the negative log-likelihood (NLL) is
\begin{align}
  \mathcal{L}_{\mathrm{NLL}}(\theta) = - \frac{1}{N_{\mathrm{shots}}} \sum_i n_i \log p_{x_i}(\theta),
\end{align}
where $p_x(\theta) = \mathrm{Tr} \left[ \rho(\theta) \dyad{x} \right]$ is computed by contracting the MPDO along the physical indices specified by bitstring $x$.
This contraction is fully differentiable as mentioned in the previous section.

An alternative to NLL is an entropy-regularized optimal transport (OT) strategy~\cite{cuturi2013sinkhorn}.
While NLL treats each bitstring independently---assigning zero credit to a simulated bitstring that differs from any observed one by even a single bit---OT exploits the distance between bitstrings so that ``near-miss'' outcomes still contribute to the gradient.
The OT-based approach can be advantageous when the number of shots is limited or the circuit is highly noisy,
as it can extract gradient information from near-miss bitstrings that NLL ignores.

Let $\mu(\theta) = \{p_x(\theta)\}$ denote the simulated distribution over bitstrings and let $\nu$ denote the empirical distribution from hardware.
A transport plan $\pi$ is a joint distribution over pairs of bitstrings $(x, y)$ whose marginals recover the two input distributions:
$\sum_y \pi_{xy} = \mu_x$ and $\sum_x \pi_{xy} = \nu_y$.
Intuitively, $\pi_{xy}$ specifies how much probability mass is moved from simulated outcome $x$ to observed outcome $y$.
The cost of this transport is measured by a ground metric;
we use the Hamming distance $C_{xy} = \sum_{i=1}^{N} |x_i - y_i|$, \textit{i.e.} the number of bit positions at which $x$ and $y$ differ---a natural choice for bitstrings from quantum circuits.
The entropy-regularized OT cost is
\begin{align}
  \label{eq:ot}
  W_\varepsilon(\mu, \nu) = \min_{\pi}
  \sum_{x,y} \pi_{xy} C_{xy}
  + \varepsilon \sum_{x,y} \pi_{xy} \log \pi_{xy},
\end{align}
where the minimization is over all valid transport plans $\pi$,
and $\varepsilon > 0$ controls the strength of entropic regularization.
The regularization term makes the optimization strictly convex,
so the minimum is unique and can be found efficiently by the Sinkhorn algorithm~\cite{cuturi2013sinkhorn}---an iterative matrix-scaling procedure that converges in $O(1/\varepsilon)$ iterations.
Crucially, each Sinkhorn iteration involves only matrix multiplications and element-wise operations, making the entire procedure differentiable.
The OT loss is
\begin{align}
  \mathcal{L}_{\mathrm{OT}}(\theta)
  = W_\varepsilon \left( \mu(\theta), \nu\right),
\end{align}
and its gradient $\nabla_\theta \mathcal{L}_{\mathrm{OT}}$ is obtained end-to-end~\footnote{
  An alternative is the REINFORCE estimator~\cite{Williams:1992mfq},
  which avoids computing exact probabilities.
  We found the direct Sinkhorn approach is more stable in practice.
}.

The noise parameters are optimized using the AdamW~\cite{loshchilov2018decoupled} optimizer implemented in Optax~\cite{deepmind}.
The entire pipeline---unitary construction via matrix exponentiation,
Kraus operator extraction,
noise channel application (including SVD truncation),
and MPDO probability computation---is implemented in JAX~\cite{jax},
so that $\nabla_\theta \mathcal{L}$ is computed via jax.value\_and\_grad in a single end-to-end pass (see also ref.~\cite{Liao:2019bye} for automatic differentiation on tensor networks).

\section{Results}
\label{sec:results}

\subsection{Experimental setup}

We train the noise model on the adder\_n10 circuit from QASMBench~\cite{Li:2020fmi},
a 10-qubit quantum ripple-carry adder based on the construction in ref.~\cite{Cuccaro:2004xxx}.
The circuit computes the addition $1 + 15 = 16$, encoding the result in a 5-bit output register with the ideal outcome $\ket{10000}$.
The circuit is transpiled into the native gate set ($\sqrt{X}$, $R_Z$, $X$, $CZ$) of ibm\_fez, a 156-qubit Heron-generation processor arranged on a heavy-hex lattice, using Qiskit's generate\_preset\_pass\_manager function with optimization level~3~\cite{qiskit2024},
resulting in a transpiled circuit of depth 344 with 210 $\sqrt{X}$ gates, 182 $R_Z$ gates, 114 $CZ$ gates, and 2 $X$ gates.
The crosstalk neighbors used by the noise model (see sec.~\ref{sec:noise_arch}) are taken from the heavy-hex coupling map of the device.
The count distribution used for training is obtained from a 16,384-shot execution on ibm\_fez.

The noise model uses $N_{\mathrm{K}} = 4$ Kraus operators for all channels.
The total parameter count is
\begin{align}
  9 \times (2N_{\mathrm{K}}^{(1)})^2 + (4N_{\mathrm{K}}^{(2)})^2 = 9 \times 64 + 256 = 832,
\end{align}
where $N_{\mathrm{K}}^{(1)} = N_{\mathrm{K}}^{(2)} = 4$ are the number of Kraus operators for single- and two-qubit channels, respectively.
Note that there are 9 single-qubit noise channels ($\mathcal{E}_{\mathrm{prep}}$, $\mathcal{E}_{\mathrm{meas}}$, $\mathcal{E}_{\sqrt{X}}$, $\mathcal{E}_{R_{Z}}$, $\mathcal{E}_{X}$, $\mathcal{E}_{\mathrm{ct}, \sqrt{X}}$, $\mathcal{E}_{\mathrm{ct}, R_{Z}}$, $\mathcal{E}_{\mathrm{ct}, X}$, $\mathcal{E}_{\mathrm{ct}, CZ}$) and only 1 two-qubit channel ($\mathcal{E}_{\mathrm{CZ}}$).

We optimize the parameters using the AdamW optimizer with learning rate 0.001 in every optimization run in this section.
For the initial Kraus operators, we use the Stinespring parameterization with all-zero entries,
which yields identity channels.
This choice is natural under the assumption that the device is well calibrated and that the true noise is close to zero.
Of course, starting from randomly generated Kraus operators is another option.
Unless otherwise noted, we use $\chi = 8$ (with $\kappa = 2\chi = 16$) for the MPDO simulation in this section.

\subsection{Optimization history}
\label{sec:history}

Figure~\ref{fig:optimization_history} shows the convergence of $\mathcal{L}_{\mathrm{NLL}}$ and $\mathcal{L}_{\mathrm{OT}}$ in training on the distribution of the adder\_n10 circuit on ibm\_fez.
Both runs use the same MPDO configuration ($\chi = 4$, $\kappa = 8$).
For the OT run, we set the entropic regularization to $\varepsilon = 0.1$ and solve with 1024 Sinkhorn iterations.
Overall, both losses show similar convergence behavior,
especially in the early and late iterations,
so for the present dataset the two strategies perform comparably.
Thus we use NLL, which shows monotonic convergence in fig.~\ref{fig:optimization_history}, exclusively in the remaining experiments;
however, the OT loss may offer an advantage in more challenging regimes---fewer shots or deeper, noisier circuits---where the Hamming metric can extract gradient information from near-miss bitstrings that NLL treats as uninformative.

\begin{figure}[htbp]
  \centering
  \includegraphics[width=\hsize]{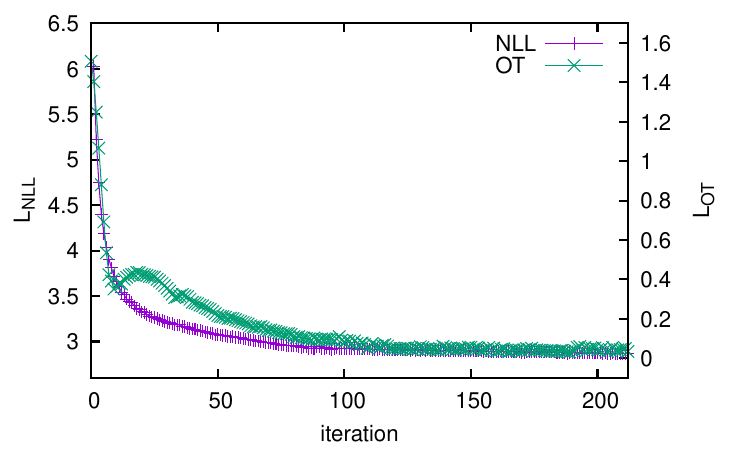}
  \caption{
    Optimization history for $\mathcal{L}_{\mathrm{NLL}}$ and $\mathcal{L}_{\mathrm{OT}}$.
    Note that the two loss scales differ, and that the values themselves are not compared to each other.
  }
  \label{fig:optimization_history}
\end{figure}

Since $\mathcal{L}_{\mathrm{NLL}}$ equals the cross-entropy $H(q, p(\theta)) = H(q) + D_{\mathrm{KL}}(q | p(\theta))$,
where $D_{\mathrm{KL}}$ is the Kullback--Leibler divergence,
its theoretical minimum is the Shannon entropy of the given distribution, $H(q) \approx 2.81$ (in nats) for this dataset.
This confirms the convergence quality of NLL.
Similarly, the OT loss converges toward zero.

\subsection{Synthetic noise}

Before digging into the real-hardware results,
we first validate our approach on synthetic data where the true noise channels are known.

We use the same adder\_n10 circuit (transpiled depth 344) and simulate it with Qulacs~\cite{Suzuki2021qulacsfast},
injecting bit-flip noise after every single-qubit gate and two-qubit depolarizing noise after every $CZ$ gate;
no noise is applied to state preparation, measurement, or neighboring qubits (crosstalk).
The bit-flip channel on a single qubit has Kraus operators
\begin{align}
  K_0 = \sqrt{1-p} I, && K_1 = \sqrt{p} X,
\end{align}
and the two-qubit depolarizing channel has
\begin{align}
  K_0 = \sqrt{1-p} I_4, && K_j = \sqrt{\frac{p}{15}} P_j \quad (j = 1,\ldots,15),
\end{align}
where $\{P_j\}$ are the 15 non-identity two-qubit Pauli operators $\{I,X,Y,Z\}^{\otimes 2} \setminus \{I \otimes I\}$.
For both channels we set $p=0.001$ in this experiment.
We collect $16,384$ shots from a full statevector simulation with Qulacs and train the model on the resulting distribution.

To quantify how well the learned channels recover the true noise, we compare them via the Choi--Jamio{\l}kowski representation~\cite{Choi:1975nug,Jamiolkowski:1972pzh},
which encodes the full action of a channel as a single $d^2 \times d^2$ matrix on a doubled Hilbert space.
For a channel $\mathcal{E}$ with Kraus operators $\{K_k\}$, the Choi matrix is
\begin{align}
  J(\mathcal{E}) = \sum_{k=0}^{N_{\mathrm{K}} - 1} \operatorname{vec}(K_k) \operatorname{vec}(K_k)^\dagger,
\end{align}
where $\operatorname{vec}(K)$ denotes column-stacking vectorization of $K$.
The normalized Choi matrix $J / \operatorname{Tr}(J)$ is a valid density matrix for trace-preserving channels.

We report two complementary metrics.
Let $\rho$ and $\sigma$ denote the normalized Choi states of the two channels being compared.
The process fidelity is the Uhlmann fidelity~\cite{Uhlmann:1976def}
\begin{align}
  \label{eq:process_fidelity}
  F(\rho, \sigma)
  = \left( \operatorname{Tr} \sqrt{\sqrt{\rho} \sigma \sqrt{\rho}} \right)^2,
\end{align}
which equals 1 if and only if the two channels are identical.
The trace distance~\cite{Fuchs:1997ss}
\begin{align}
  \label{eq:trace_distance}
  T(\rho, \sigma)
  = \frac{1}{2} \left\| \rho - \sigma \right\|_1
\end{align}
quantifies the maximum probability of distinguishing the two channels and is bounded by the Fuchs--van de Graaf inequality $1 - \sqrt{F} \leq T \leq \sqrt{1 - F}$.
Both metrics are independent of the particular Kraus decomposition, since the Choi matrix is unique for a given channel.

Table~\ref{tab:sim_fidelity} shows the process fidelity and trace distance as a function of the MPDO bond dimension $\chi$ (with inner dimension $\kappa = 2\chi$).
We observed that the NLL loss decreases monotonically (1.75, 1.40, 1.27) with increasing $\chi$,
reflecting the improved representational capacity of the MPDO.
Since the single-qubit channels do not involve site-site connections, they are learned successfully even at the smallest possible bond dimension $\chi=2$.
On the other hand, the two-qubit error channel for the $CZ$ gate is less accurately recovered, with $F = 0.79$.
This is expected, as two-qubit noise generates entanglement between sites that requires a larger bond dimension to represent faithfully.
Increasing the bond dimension to $\chi = 4$ and $8$ brings all channels above $F = 0.99$.

This recovery of the ground-truth channels is not automatic from distribution matching alone.
The training objective only requires the simulated distribution to match the observed one on this specific circuit,
and in general many different noise models can produce the same output distribution on a fixed circuit.
The fact that our learned channels nevertheless converge close to the ground truth in the Choi fidelity provides empirical support for the legitimacy of our modeling choices.
We attribute this to a combination of factors:
(i)~the deep, non-Clifford gate composition (arbitrary-angle $R_Z$ rotations) of adder\_n10 exercises each gate type many times on diverse input states, so a single shared channel per gate type is tightly constrained;
(ii)~the Stinespring parameterization (CPTP by construction) and near-identity initialization confine the optimizer to a small neighborhood of the true channels;
(iii)~in the weak-noise regime, output-distribution proximity and Choi fidelity can happen to align, so distribution-matching translates to channel recovery.
We emphasize that this identifiability is circumstantial rather than guaranteed---shorter or more symmetric circuits or stronger noise regimes may break the correspondence.

Moreover, the success in learning the noise on $X$ is surprising,
since there are only 2 $X$ gates in the original and transpiled adder\_n10 circuit.
As discussed just above, a possible reason is that the near-identity ground truth, the identity initialization, and the 16,384 shots of training data make even a rare gate learnable.

\begin{table}[htbp]
  \centering
  \caption{
    Process fidelity and trace distance between learned and true noise channels for the simulated adder\_n10 circuit with bit-flip ($p = 0.001$) on single-qubit gates and depolarizing ($p = 0.001$) on $CZ$ gates, as a function of MPDO bond dimension $\chi$ (inner dimension is always $\kappa = 2\chi$).
  }
  \label{tab:sim_fidelity}
  \begin{tabular}{clcc}
    \hline\hline
    $\chi$ & Channel & Proc.\ fidelity & Trace dist.\ \\
    \hline
    2 & $\sqrt{X}$ (bit-flip) & 0.9889 & 0.1002 \\
           & $R_Z$ (bit-flip)      & 0.9882 & 0.104 \\
           & $X$ (bit-flip)        & 0.9524 & 0.2066 \\
           & $CZ$ (depolarizing)   & 0.7918 & 0.29 \\
    \hline
    4 & $\sqrt{X}$ (bit-flip) & 0.9978 & 0.02939 \\
           & $R_Z$ (bit-flip)      & 0.9972 & 0.03052 \\
           & $X$ (bit-flip)        & 0.9988 & 0.02147 \\
           & $CZ$ (depolarizing)   & 0.9958 & 0.04584 \\
    \hline
    8 & $\sqrt{X}$ (bit-flip) & 0.9991 & 0.01363 \\
           & $R_Z$ (bit-flip)      & 0.9995 & 0.01423 \\
           & $X$ (bit-flip)        & 0.9984 & 0.008822 \\
           & $CZ$ (depolarizing)   & 0.9982 & 0.02072 \\
    \hline\hline
  \end{tabular}
\end{table}

\subsection{IBM hardware noise}
\label{sec:ibm_hardware_noise}

Figure~\ref{fig:dist_adder} overlays the ibm\_fez measurement distribution for adder\_n10 with samples drawn from the MPDO after training the noise model on the same hardware data.
The correct output $\ket{10000}$ is the most frequently observed bitstring on the device ($\approx 30\%$ of shots), but the deep transpiled circuit with depth 344 spreads probability across all $2^5 = 32$ outcomes in a long-tailed distribution.
The learned model reproduces the overall shape of the hardware distribution:
the correct bitstring dominates at a comparable probability, and the ranking and relative magnitudes of the remaining bitstrings closely match the experimental data.
Quantitatively, the final NLL of the learned model on this distribution is $\approx 2.87$, close to the Shannon entropy lower bound $\approx 2.81$ as discussed in sec.~\ref{sec:history}.
This demonstrates that the differentiable Kraus operator framework can learn an effective noise model that reproduces the output distribution of a real quantum device executing a moderately deep circuit.

\begin{figure}[htbp]
  \centering
  \includegraphics[width=\hsize]{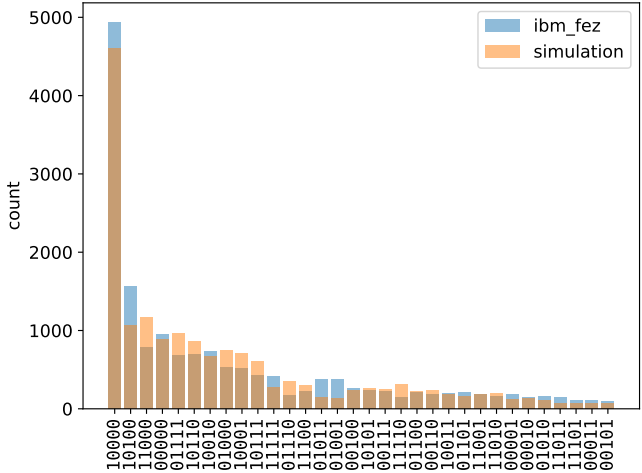}
  \caption{
    Measurement distribution overlay for the adder\_n10 circuit on ibm\_fez (16,384 shots).
    Bars show device counts and samples from the MPDO with Kraus operators learned on the same device data.
    The ideal output $\ket{10000}$ is the most probable outcome, and the learned model reproduces both the dominant peak and the noise-induced tail across all 32 bitstrings.
  }
  \label{fig:dist_adder}
\end{figure}

A learned noise model is most useful when it generalizes beyond the circuit on which it was trained.
To test this, we apply the Kraus parameters learned from the adder\_n10 hardware data without retraining to a different QASMBench circuit, multiply\_n13, a 13-qubit quantum integer multiplier that computes $3 \times 5 = 15$ with ideal outcome $\ket{1111}$, with a different gate composition ($\sqrt{X}$: 178, $R_{Z}$: 122, $CZ$: 90, $X$: 6) and a different transpiled depth 175.
Figure~\ref{fig:dist_multiply} overlays the ibm\_fez device counts for multiply\_n13 with MPDO samples generated using the adder\_n10-trained Kraus parameters.
The learned model continues to track the shape of the device distribution on this unseen circuit, indicating that the effective noise it captures is largely circuit-independent and reflects intrinsic device characteristics rather than overfitting to the training circuit.
To confirm that this agreement is non-trivial, we also simulated multiply\_n13 with a random noise model whose Stinespring parameters are drawn from $\mathcal{N}(0, \sigma^2)$ with $\sigma$ set to match the root-mean-square (RMS) magnitude of the learned parameters across all channels.
As the inset of fig.~\ref{fig:dist_multiply} shows, the random baseline produces a quasi-flat distribution, demonstrating that the learned channels encode device-specific structure beyond what generic CPTP noise of comparable strength can provide.

\begin{figure}[htbp]
  \centering
  \includegraphics[width=\hsize]{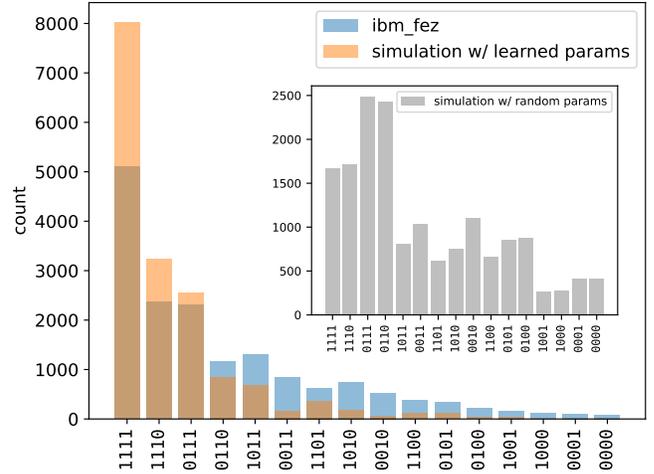}
  \caption{
    Generalization test: device vs. MPDO-simulated distributions for multiply\_n13 on ibm\_fez.
    The main plot overlays the device counts with samples from the MPDO using the Kraus operators learned from the adder\_n10 hardware data, without retraining for this circuit.
    The learned noise model successfully tracks the device distribution on a circuit it has not seen during training.
    The inset shows the distribution produced by a random noise model with matched parameter magnitude;
    the random baseline fails to capture the device distribution.
  }
  \label{fig:dist_multiply}
\end{figure}

To quantify generalization more systematically, we evaluate the learned noise model on $\sim 40$ circuits from QASMBench, none of which are used for training.
For each circuit, we compute the classical fidelity $F(p, q) = \left(\sum_x \sqrt{p_x q_x}\right)^2$ between the device distribution $p$ and the simulated distribution $q$.
As a baseline, we also simulate each circuit with random CPTP noise of matched parameter magnitude, generated in the same manner as the random baseline used for multiply\_n13 above.
Figure~\ref{fig:classic_fidelity} plots the fidelity of the learned model against that of the random model for each circuit;
the majority of points lie above the diagonal, indicating that the learned model systematically outperforms the random baseline.
The few cases where the random baseline achieves comparable or higher fidelity correspond to circuits whose device output distribution is nearly uniform or random, lacking dominant peaks or clear outcome ranking that the learned noise structure could improve upon.
We note that the random baseline has a mild information leakage: its overall noise magnitude is set by the RMS of the learned parameters.

\begin{figure}[htbp]
  \centering
  \includegraphics[width=\hsize]{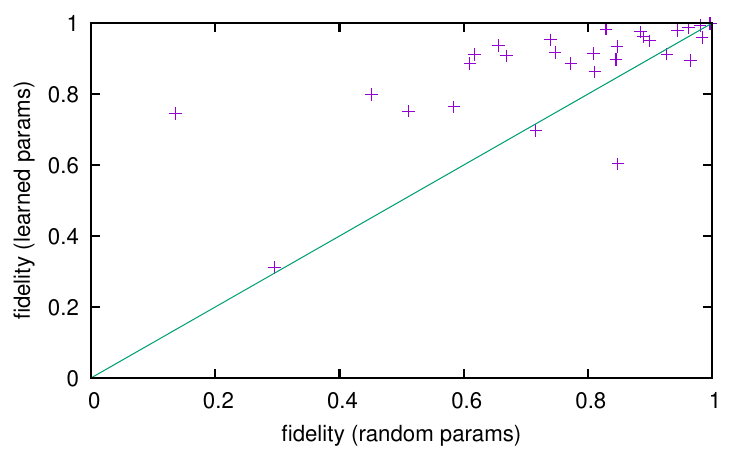}
  \caption{
    Classical fidelity between device and simulated output distributions for QASMBench circuits not used for training.
    Each point compares the fidelity achieved by the learned noise model (vertical axis) against a random CPTP noise model of matched parameter magnitude (horizontal axis).
    Points above the diagonal indicate that the learned model better reproduces the device distribution.
  }
  \label{fig:classic_fidelity}
\end{figure}

The learned Kraus parameters allow us to quantify how close each individual channel is to its noiseless ideal.
Each learned channel $\mathcal{E}$ models the noise that is composed with the ideal operation (gate, crosstalk, prep, or measurement),
so the fidelity of the noisy operation against its ideal equals the fidelity of $\mathcal{E}$ against the
identity channel $\mathcal{I}$.
In this special case, the Choi state of $\mathcal{I}$ is the pure maximally entangled state $\dyad{\Phi^{+}}$,
and eq.~\eqref{eq:process_fidelity} reduces to the simplified entanglement fidelity
\begin{align}
  \label{eq:process_fidelity_id}
  F(\rho, \dyad{\Phi^{+}})
  = \expval{\rho}{\Phi^{+}}
  = \frac{1}{d^2} \sum_{k=0}^{N_{\mathrm{K}}-1} \left| \operatorname{Tr}(K_k) \right|^{2},
\end{align}
which avoids any matrix square root.
We report the average gate fidelity
$F_{\mathrm{avg}} = (d F + 1) / (d + 1)$,
which is the standard hardware-spec figure of merit.
Table~\ref{tab:ibm_fez_infidelity} shows the per-channel infidelity $1 - F_\text{avg}$ for the noise model learned on ibm\_fez.
The dominant source of error is the two-qubit $CZ$ channel ($\approx 2.03 \times 10^{-2}$).
The error on the single-qubit $X$ gate is also relatively large ($\approx 1.22 \times 10^{-2}$),
although the impact on the adder\_n10 and multiply\_n13 circuits would be small owing to the fewer $X$ gates.
The magnitudes of all the other noise channels sit at the $10^{-5}$--$10^{-3}$ level.

\begin{table}[htbp]
  \centering
  \caption{
    Per-channel infidelity $1 - F_{\mathrm{avg}}$ extracted from the learned Kraus parameters for ibm\_fez.
    $N_{\mathrm{K}}=4$.
    For the crosstalk entries, $1 - F_{\mathrm{avg}}$ is the average error induced on each nearest-neighbor qubit per nearby gate.
  }
  \label{tab:ibm_fez_infidelity}
  \begin{tabular}{lcc}
    \hline\hline
    Channel & $d$ & $1 - F_\text{avg}$ \\
    \hline
    Prep & 2 & $3.9\times10^{-4}$ \\
    Meas & 2 & $1.28\times10^{-3}$ \\
    \hline
    $\sqrt{X}$ & 2 & $7.3\times10^{-5}$ \\
    $R_Z$      & 2 & $7.07\times10^{-5}$ \\
    $X$        & 2 & $1.22\times10^{-2}$ \\
    $CZ$       & 4 & $2.03\times10^{-2}$ \\
    \hline
    $\sqrt{X}$ crosstalk & 2 & $7.39\times10^{-5}$ \\
    $R_Z$ crosstalk      & 2 & $7.5\times10^{-5}$ \\
    $X$ crosstalk        & 2 & $3.43\times10^{-3}$ \\
    $CZ$ crosstalk       & 2 & $7.33\times10^{-5}$ \\
    \hline\hline
  \end{tabular}
\end{table}

These per-channel numbers should be interpreted as effective error budgets rather than faithful reproductions of the physical gate errors on ibm\_fez.
Several modeling choices could inflate the learned $CZ$ infidelity above the officially reported value ($\approx 2.54 \times 10^{-3}$ median on Apr. 15, 2026~\cite{IBMQuantum}).
First, our noise model is deliberately simplified, assigning a single shared channel to each gate type and treating preparation, measurement, and crosstalk as independent single-qubit channels (see sec.~\ref{sec:noise_arch}).
The training objective only requires matching the output distribution,
so any residual error unaccounted for by the model is spread across the channels.
Second, we use $N_{\mathrm{K}} = 4$ Kraus operators for the $CZ$ channel, below the $d^2 = 16$ needed to represent an arbitrary two-qubit CPTP map.
This forces the learned effective $CZ$ channel to approximate the underlying noise within a restricted manifold,
adding a representation-error contribution.
The learned per-channel infidelities thus serve as effective upper bounds on per-gate error contributions, with some channels acting as sinks for error sources that the simplified architecture cannot otherwise attribute.

\subsection{Application: noisy QAOA simulation for LABS problem with error detection}
\label{sec:labs}

The low autocorrelation binary sequences (LABS) problem~\cite{1053969,1054411} asks for a sequence $s \in \{-1, +1\}^N$ that minimizes the sum of squared aperiodic autocorrelations,
\begin{align}
  \label{eq:labs_energy}
  E(s) = \sum_{k=1}^{N-1} C_k(s)^2, && C_k(s) = \sum_{i=1}^{N-k} s_i s_{i+k},
\end{align}
or equivalently to maximize the merit factor $\mathcal{F}(s) = N^2 / (2 E(s))$.
LABS arises \textit{e.g.} in signal-design problems for radar/sonar, and is NP-hard;
the optimal sequences are known only up to $N \approx 66$ and heuristics are relied upon beyond that.
Encoding $s_i \to (1 - 2 z_i)$ with $z_i \in \{0,1\}$ maps $E(s)$ onto an Ising Hamiltonian containing both two-body $ZZ$ and four-body $ZZZZ$ interactions,
giving a dense and long-range cost Hamiltonian on $N$ qubits.
In ref.~\cite{Shaydulin:2023fpr}, the quantum approximate optimization algorithm (QAOA)~\cite{Farhi:2014ych}\cite{Blekos:2023nil} with an error detection scheme was applied to this problem.
Their scheme was implemented on Quantinuum trapped-ion quantum processors~\cite{Pino:2020mku,Moses:2023ozv} up to $N=18$,
making it a natural target for noise-aware simulations.

To illustrate how the learned channels can be put to practical use in the noisy intermediate-scale quantum--early fault-tolerant quantum computing era,
we apply the noise model trained on ibm\_fez to single-layer LABS QAOA for several values of $N$,
using the optimal QAOA parameters reported in ref.~\cite{Shaydulin:2023fpr}.
Since the major source of error in QAOA circuits is the phase separator layer that involves multi-qubit gates,
we simply assume that only this layer is noisy, by setting all channels other than $CZ$ to identity.
We wrap the phase separator in a parity-check error-detection scheme that uses two ancilla qubits to flag $X$- and $Z$-type errors that occur during the phase-separator block;
only samples for which both ancillas measure $0$ are accepted~\footnote{
  Note that, in ref.~\cite{Shaydulin:2023fpr}, the error-detection scheme was threefold.
}.
In this experiment, we further simplify the situation by suppressing noise channels that act on the error-detection part.

Figure~\ref{fig:parity_check} shows the circuit to be simulated.
The LABS cost Hamiltonian commutes with both $\bigotimes_i Z_i$ and $\bigotimes_i X_i$,
so its time evolution preserves the global $Z$- and $X$-parities.
By mapping these parities onto the two ancillas before the phase separator with a $CZ$ chain for the $Z$-parity and a CNOT chain for the $X$-parity,
a phase-separator error that flips a global parity can be detected by measuring the ancillas.
Both chains are undone after the phase separator layer.

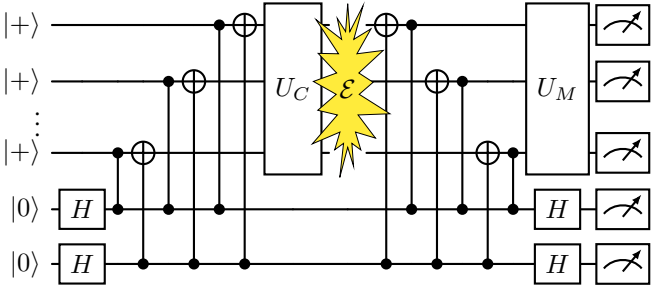
\begin{figure}[htbp]
  \centering
  \tikzset{
    noisy/.style={starburst, fill=yellow, line width=0pt, inner xsep=-4pt, inner ysep=-16pt},
  }
  \begin{quantikz}[column sep=0.1cm, row sep=0.2cm,
    wire types={q,q,n,q,q,q}]
    \lstick{$\ket{+}$} &          &            &           &           &           & \control{}& \targ{}   & \gate[4]{U_C} & \gate[4,style={noisy}]{\mathcal{E}} & \targ{}    & \control{}  &           &           &         &        & \gate[4]{U_M} & \meter{} \\
    \lstick{$\ket{+}$} &          &            &           & \control{}& \targ{}   &           &           &               &                        &           &           & \targ{}   & \control{}&           &            &               & \meter{} \\
    \lstick{$\vdots$}  &          &            &           &           &           &           &           &               &                        &           &           &           &           &           &            &               & \\
    \lstick{$\ket{+}$} &          & \control{} & \targ{}   &           &           &           &           &               &                        &           &           &           &           &  \targ{}  & \control{} &               & \meter{} \\
    \lstick{$\ket{0}$} & \gate{H} & \ctrl{-1}  &           & \ctrl{-3} &           & \ctrl{-4} &           &               &                        &           & \ctrl{-4} &           & \ctrl{-3} &           & \ctrl{-1}  & \gate{H}      & \meter{} \\
    \lstick{$\ket{0}$} & \gate{H} &            & \ctrl{-2} &           & \ctrl{-4} &           & \ctrl{-5} &               &                        & \ctrl{-5} &           & \ctrl{-4} &           & \ctrl{-2} &            & \gate{H}      & \meter{} \\
  \end{quantikz}
  \caption{
    Single-layer QAOA circuit with parity-check error detection.
    The two bottom wires are ancillas that encode the global $Z$-parity (via $CZ$ to each data qubit) and the global $X$-parity (via CNOT to each data qubit).
    The phase separator $U_{\mathrm{C}} = e^{-i\gamma H_{\mathrm{C}}}$ is applied to the data qubits, with the learned ibm\_fez noise acting on its constituent $CZ$ gates.
    The parity encoding is then undone,
    the mixer $U_{\mathrm{M}} = e^{-i\beta \sum_j X_j}$ is applied,
    and all qubits are measured.
    All operations except $CZ$ in $U_{\mathrm{C}}$ are treated as noiseless.
    Samples are accepted only when both ancillas read 0;
    phase-separator error that flips a global parity is rejected.
  }
  \label{fig:parity_check}
\end{figure}

Figure~\ref{fig:labs_acceptance} shows the post-selection acceptance rate as a function of $N$,
and fig.~\ref{fig:labs_mf} shows the expectation value of the merit factor over the accepted samples.
16,384 samples are taken from the MPDO simulator with $\chi = 32$ (convergence in $\chi$ was confirmed) for each $N$.
As $N$ grows, the acceptance rate decreases (more two-qubit gates accumulate more detectable errors).
In this range, the results with error-detection approach the noiseless simulation results almost monotonically, in contrast to those without error-detection.
Note, however, that the noisy simulation results converge to the random-sampling result regardless of whether error-detection is used.
We confirmed that the results for $N \geq 8$ all behave similarly.
This means that $N = 8$ is already too difficult given the noise strength.
A rough estimate supports this:
for $N = 8$, the transpiled phase separator contains 268 $CZ$ gates,
so with the learned $CZ$ infidelity of $\approx 2 \times 10^{-2}$ the expected number of $CZ$ errors per shot is $\approx 268 \times 0.02 \approx 5$,
making the probability of an error-free shot only $\approx 0.98^{268} \approx 0.5\%$.
Moreover, parity-based post-selection has an intrinsic blind spot:
any event in which an even number of parity-breaking errors occur leaves the global $Z$- and $X$-parities unchanged and is accepted as a clean sample.
Unlike full error correction, the scheme cannot distinguish an error-free shot from one with an even number of parity-breaking errors,
and this limitation becomes increasingly relevant as the expected error count per shot grows.

The demonstration in this section is exactly the type of analysis that the learned noise model enables:
offline circuit-level feasibility assessment of quantum algorithms under realistic device-level noise, using only a single calibration circuit as training data.

\begin{figure}[htbp]
  \centering
  \includegraphics[width=\hsize]{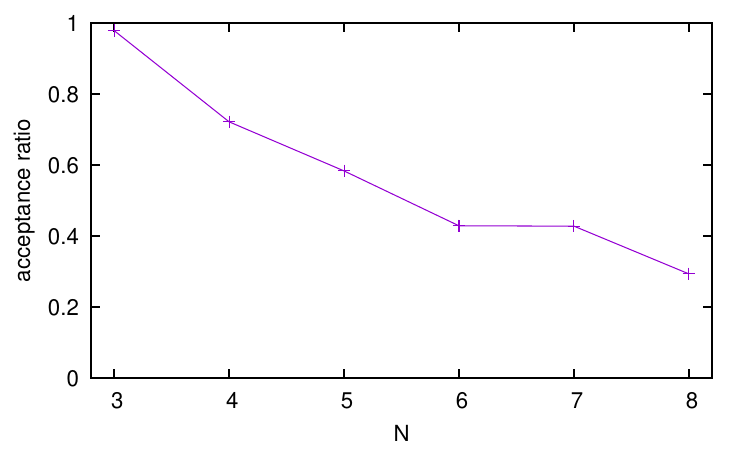}
  \caption{
    Post-selection acceptance rate of the parity-check error-detection scheme.
  }
  \label{fig:labs_acceptance}
\end{figure}

\begin{figure}[htbp]
  \centering
  \includegraphics[width=\hsize]{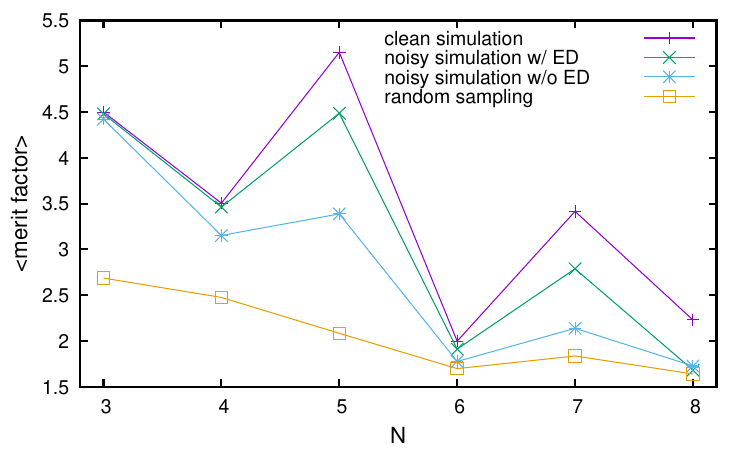}
  \caption{
    Expectation value of the merit factor for the accepted LABS QAOA samples with and without error-detection.
    For comparison, noiseless (clean) simulation results and random sampling results are also shown.
  }
  \label{fig:labs_mf}
\end{figure}

\section{Summary}
\label{sec:summary}

We have presented a method for learning quantum hardware noise as automatically differentiable Kraus operators and simulating the resulting noisy circuits on MPDO~\cite{Verstraete:2004gdw,2021PhRvR...3b3005C,Guo:2023dqn}.
The key ingredient on the channel side is a Stinespring-based parameterization in which a real parameter vector $\theta$ is mapped to a Hermitian matrix $H(\theta)$,
then to a dilated unitary $U(\theta) = e^{i H(\theta)}$, from which the Kraus operators are extracted by block slicing.
This construction is CPTP by construction for any $\theta$, is surjective onto the set of CPTP maps with at most $N_{\mathrm{K}}$ Kraus operators, and is differentiable,
so the entire pipeline---parameters, Kraus operators, MPDO simulation, and measurement-probability evaluation---can be trained end-to-end with gradient-based optimization in JAX and Optax.
On top of this channel parameterization, we built a noise-model architecture that attaches independent learnable channels to each native gate type, every nearest-neighbor qubit affected by crosstalk, and state preparation and measurement,
with the simplifying assumption that preparation, measurement, and crosstalk channels remain single-qubit.

The forward model is an MPDO, whose explicit inner (Kraus) index preserves Hermiticity and positivity under bond truncation---a property that plain-MPO density-matrix representations (\textit{e.g.} \cite{Lee:2025jpe}) do not share.
Noise parameters are optimized against a distribution from hardware with either an NLL or an entropy-regularized OT loss using the Hamming ground metric.

On synthetic data with known bit-flip and two-qubit depolarizing noise injected into the transpiled adder\_n10 circuit,
the learned channels converged close to the ground truth in terms of the Choi-state process fidelity, reaching $F > 0.99$ for all channels,
including the two-qubit $CZ$ channel that requires a larger bond dimension to represent entangling noise faithfully.

On ibm\_fez hardware data, a noise model trained on adder\_n10 reproduced the distribution of the device,
including the dominant peak and the ranking of the outcomes.
More importantly, the same learned parameters, applied without retraining to the unrelated multiply\_n13 circuit, continued to track the device distribution,
indicating that what we learned is largely an intrinsic device characteristic rather than an overfit to the training circuit;
indeed, a systematic evaluation across a range of QASMBench circuits confirms that the learned model consistently outperforms random noise of matched magnitude in classical fidelity.
The per-channel infidelities isolate $CZ$ ($\approx 2 \times 10^{-2}$) and $X$ ($\approx 10^{-2}$) as the dominant error sources, with all other channels at the $10^{-5}$--$10^{-3}$ level;
these numbers should be read as effective error budgets of the simplified model with $N_{\mathrm{K}} = 4 < 16$ on $CZ$,
not as calibration-grade gate errors, and accordingly act as upper bounds that absorb unattributed error.

As an application, we used the ibm\_fez-trained noise model to simulate single-layer QAOA for the LABS problem~\cite{1053969,1054411} wrapped in a parity-check error-detection scheme~\cite{Shaydulin:2023fpr},
applying the learned noisy channel only to the phase separator part.
The merit factor of the post-selected samples tends to move monotonically toward the noiseless value.
This is precisely the kind of offline, circuit-level feasibility assessment that the learned noise model is meant to enable:
starting from a single calibration circuit experiment on real hardware,
we can predict how a different algorithm of interest will behave under realistic device noise without returning to the device.

Relative to the prior works surveyed in sec.~\ref{sec:introduction},
our method occupies a distinctive corner of the design space:
we learn a full, device-wide, circuit-agnostic CPTP noise model end-to-end through an MPDO forward model,
from computational-basis counts of a single circuit experiment on real hardware,
and we demonstrate that the learned model transfers across circuits.

We emphasize that the $\leq 13$-qubit ceiling in our experiments reflects the noise strength of current hardware rather than any intrinsic limit of the method:
with the per-$CZ$ infidelity of IBM Heron-generation devices, circuits beyond this size already approach the random-sampling regime and no longer offer a meaningful learning target.
The MPDO forward model and the Stinespring parameterization themselves scale to substantially larger systems,
and we expect the same pipeline to remain directly applicable as device error rates improve.

Several directions for future work naturally follow from these observations.
First, other hardware platforms are attractive targets for the same pipeline.
Trapped-ion processors, for instance, have a markedly different crosstalk signature from superconducting devices---long-range, all-to-all rather than nearest-neighbor on a heavy-hex lattice---so the crosstalk component of our noise-model architecture would need to be adjusted accordingly,
although the channel-side Stinespring parameterization and the MPDO forward model carry over unchanged.
Second, our noise architecture is deliberately simple:
a single shared channel per gate type,
and---most notably---preparation, measurement, and crosstalk all modeled as independent single-qubit channels even though the underlying processes are physically multi-qubit (correlated readout, joint system-neighbor dynamics during a gate).
Promoting these to genuine two-qubit (or larger) channels, together with more device-specific refinements, is a straightforward extension within the same framework and is worth exploring when enough training data is available to keep overfitting in check.
Third, and perhaps most fundamentally, our channels are purely Markovian and time-independent by construction:
amplitude damping, dephasing, and other time- or history-dependent error processes collapse into an effective static channel that is applied identically at every position in the circuit.
A more faithful treatment would combine the differentiable Kraus parameterization with a process tensor representation of non-Markovian dynamics,
along the lines of \textit{e.g.} ref.~\cite{White:2023kpy}, allowing the learned object to capture temporally correlated noise without giving up end-to-end gradient-based training.
Fourth, the cross-circuit generalization observed in sec.~\ref{sec:ibm_hardware_noise} was obtained by training on a single calibration circuit;
a multi-objective training strategy that simultaneously fits the output distributions of several structurally different circuits should further improve generalization by exposing each channel to a more diverse range of inputs and reducing circuit-specific biases.

Another natural direction for future work is to use the learned noise model for quantum error mitigation.
Unsupervised methods such as Q-Cluster~\cite{Patil:2025wtv} already show that noise-aware post-processing of hardware output distributions can recover dominant clean outcomes without any parametric noise model;
a learned per-channel description of the device noise should enable more targeted, model-driven mitigation strategies---\textit{e.g.} inverting specific learned channels, reweighting samples by their estimated likelihood under the clean versus noisy models, or feeding the learned Kraus operators into probabilistic error cancellation~\cite{Temme:2016vkz,Berg:2022ugn}---using the same single-circuit calibration pipeline demonstrated here.

\appendices

\section*{Acknowledgments}

We acknowledge the use of IBM Quantum Credits via the IBM Quantum Startups Program for this work.
The views expressed are those of the authors and do not reflect the official policy or position of IBM or the IBM Quantum Platform team.

This work was supported by computational resources of the Quantum-AI Hybrid Computing Infrastructure (ABCI-Q),
awarded under the ``ABCI-Q Grand Challenge'' Program of the National Institute of Advanced Industrial Science and Technology (AIST).

\printbibliography[title=References]

\end{document}